\documentclass[aps,twocolumn,prd]{revtex4}

\usepackage{amsmath}
\usepackage[dvips]{graphicx}
\usepackage{amssymb}
\usepackage{verbatim}
\usepackage{mathrsfs}
\usepackage{color,psfrag}

\newcommand{\alphap}{\ensuremath{\alpha^{\prime}}}

\newcommand{\mpl}{m_{\mathrm{Pl}}}

\newcommand{\ben}{\begin{equation}}
\newcommand{\een}{\end{equation}}
\newcommand{\bea}{\begin{eqnarray}}
\newcommand{\eea}{\end{eqnarray}}

\begin{document}
%%%%%%%%%%%%%%%%%%%%%%%%%%%%%%%%%%%%%%%%%%%%%%%%%%%%%%%%%%%%%%%%%%%%%%

\title{Large Radius Hagedorn Regime in String Gas Cosmology}
\date{July 14, 2008}
\author{Dimitri P. Skliros}
\email{d.p.skliros@sussex.ac.uk}
\author{Mark B. Hindmarsh}
\email{m.b.hindmarsh@sussex.ac.uk}
\affiliation{Department of Physics and Astronomy, University of Sussex, Brighton, East Sussex BN1 9QH, UK} 

\begin{abstract} 
We calculate the equation of state of a gas of strings at high density in a large toroidal universe, and use it to determine the cosmological evolution of background metric and dilaton fields in the entire large radius Hagedorn regime, $(\ln S)^{1/d}\ll R\ll S^{1/d}$ (with $S$ the total entropy). The pressure in this regime is not vanishing but of $\mathcal{O}(1)$, while the equation of state is proportional to volume, which makes our solutions significantly different from previously published approximate solutions. For example, we are able to calculate the duration of the high-density ``Hagedorn'' phase, which increases exponentially with increasing entropy, $S$. We go on to discuss the difficulties of the scenario, quantifying the problems of establishing thermal equilibrium and producing a large but not too weakly-coupled universe.
\end{abstract}

\maketitle
%%%%%%%%%%%%%%%%%%%%%%%%%%%%%%%%%%%%%%%%%%%%%%%%%%%%%%%%%%%%%%%%%%%%%%%%%
\section{Introduction}
\parindent = 2em
Strings may have played an important role in the very early universe due to the potential dominance of stringy effects at high energies. It is important therefore to understand the dynamics of an ensemble of strings in what seems to be a rather natural background, namely dilaton gravity, the stringy version of Einstein gravity.
This approach to string cosmology started with the work of Brandenberger and Vafa (BV) \cite{Brandenberger:1988aj} who laid down the main conceptual framework and subsequently Tseytlin and Vafa (TV) \cite{Tseytlin:1991xk} who introduced dilaton gravity  into the picture, hence making a first step towards realising a string cosmology at finite temperature. This framework incorporates all the degrees of freedom of perturbative string theory, namely oscillator, momentum and winding modes, which at high densities produce a state with unusual thermodynamic properties such as a limiting temperature, known as the Hagedorn phase. An early Hagedorn phase may turn out to have crucial consequences for the early universe, the initial singularity and the dimensionality of spacetime (see \cite{Brandenberger:1988aj} for further details).

Even though this scenario has been studied and extended in a number of different ways (see eg \cite{Cleaver:1994bw,Sakellariadou:1995vk,Brandenberger:2001kj,Easther:2002mi,Bassett:2003ck,Easther:2004sd,Takamizu:2006sy,Enqvist:2007sd,Borunda:2006fx,Biswas:2006bs,Arapoglu:2004yf,Kaya:2003py}), a complete string theory analysis has yet to be made. Furthermore, a viable string cosmology should also incorporate the attractive features of the standard cosmological model, such as the absence of a flatness and horizon problem and solutions to the entropy and size problems.

We shall here discuss the scenario introduced by TV \cite{Tseytlin:1991xk} which was inspired by the conceptual framework of BV \cite{Brandenberger:1988aj}. We shall concentrate on \emph{large} radius evolution, $\sqrt{\alphap}\ll R\ll S^{1/d}$ (with $S$ the total entropy), where the dominant contribution to the pressure is from momentum modes (roughly speaking the centre of mass motion of small loops) and hence not negligibly small. In particular, we shall be tracing the evolution of the entire large radius Hagedorn phase up until $R\sim S^{1/d}$ beyond which the universe is radiation dominated. The large radius regime had been poorly studied to date and only few analytic results had been obtained. Furthermore, analysing the large radius dynamics is crucial in order to understand various transitions that occur and to make contact with late time evolution (by which we mean radiation domination onwards). Finally, a large radius universe cannot be avoided in 
the dilaton gravity realisation of String Gas cosmology, making the reasonable assumption that the massless string degrees of freedom which take over at the end of the  Hagedorn phase are identified with the particle content of the Standard Cosmology. A simple calculation then shows that the compactification radius must be at least $10^{-2}$ mm in order to expand to at least the Hubble radius today.

It has been proposed that the string gas initial conditions can produce scale-free cosmological perturbations \cite{Nayeri:2005ck,Brandenberger:2006vv}, although the proposal was strongly criticised in \cite{Kaloper:2006xw}. In order to rescue the scenario more recent variants  depart from the original simplicity of the scenario by fixing the dilaton and introducing extra terms and fields \cite{Biswas:2006bs,Brandenberger:2007xu,Brandenberger:2007dh}. Another variant supposes that before the perturbative Hagedorn phase of the original scenario there was a strongly-coupled Hagedorn phase with constant dilaton \cite{Brandenberger:2006pr}.  Our analysis applies equally to the perturbative Hagedorn phase of this scenario -- this is independent of whether the dilaton is initially strongly coupled provided the low energy effective action is valid when the universe is much larger than string scale.

In section \ref{GF} we give a brief overview of the derivation of the equations of motion of string cosmology to set our notation conventions. In section \ref{SEF1} we reformulate the equations of motion so as to make them analytically solvable and discuss their general structure while making explicit the constraints that appear. We work in the string frame but the Einstein frame evolution is also briefly discussed, especially when making contact with late time evolution and observational constraints. In section \ref{ST} we derive the equation of state parameter which (to leading order) turns out to be proportional to the inverse entropy density. We then go on to derive the number density associated with winding and momentum modes of string, which correspond roughly speaking to long and short string loops respectively. We notice that at radii larger than $\sim(\ln S)^{1/d}$ the contribution of long string to the number density is negligible compared to that of small loops while at radii greater than $\sim S^{1/d}$ there is not enough energy to excite massive modes of string and only zero modes are present, i.e. radiation. In section \ref{SEF2} we solve the equations of motion analytically with evolving equation of state for the entire large radius Hagedorn regime and compare our results with the corresponding numerical solutions. We find that neglecting the evolution of the equation of state cannot be justified in the large radius Hagedorn regime and emphasise that the pressure is not negligible but is rather $\mathcal{O}(1)$ in string units and approximately constant. The problems of this scenario are discussed in section \ref{SEP}. In particular, the issues of Jeans instabilities and maintaining thermal equilibrium are examined and we improve on previous estimates \cite{Danos:2004jz,Takamizu:2006sy} by showing that the universe could not have been in thermal equilibrium at any stage of the evolution in the large radius Hagedorn region if string is weakly coupled, $g\ll1$. We also discuss the size problem, and its relation to the weak-coupling problem in String Gas cosmology.

%%%%%%%%%%%%%%%%%%%%%%%%%%%%%%%%%%%%%%%%%%%%%%%%%%%%%%%%%%%%%%%%%%%%%%%
\section{General Formalism}\label{GF}
\parindent = 2em
The critical tree level dilaton-gravity low energy effective action plus stringy matter contributions reads \footnote{We adopt the signature $-++\cdots$.} (see eg
\cite{Polchinski:1998rr})
\begin{multline}\label{eq: action full}
S = \frac{1}{2\kappa_{10}^2}\int\limits_{\mathcal{T}^9\times\mathbb{R}}
d^{10}x\sqrt{|G|}e^{-2\phi}\\
\times\left[R_{10}+4G^{AB}\nabla_A\phi\nabla_B\phi+\dots 
\right]+S_M,
\end{multline}
where $\phi$ is the dilaton, $R_{10}$ is the 10-dimensional string frame Ricci scalar and $G_{AB}$ with $A,B=0,\cdots,9$ is the string frame metric tensor \footnote{Lorentz invariance is maintained locally because the tangent spaces $T_p\mathcal{M}_{10}=T_p(\mathcal{T}^9\times\mathbb{R})$ at $p$.}. The first two terms represent the lowest order (in both $\alpha^{\prime}$ and $g_s$) contribution to the bosonic NS-NS sector and are present in all 5 superstring low energy effective actions \footnote{We have not included the antisymmetric tensor $B_{(2)}$ contribution because its contribution is just that of a massless derivatively coupled scalar field which can be dropped as it offers no new possibilities. A theory of a $p$-form gauge field is equivalent to a theory of a $(D-p-2)$-form gauge field which for $p=2$ and $D=4$ is just a scalar field (see eg \cite{Weinberg:1995mt})}.  The string gas contribution is contained in the matter action, $S_M$, the form of which is derived from the string spectrum on a torus at finite temperature. We are thus coupling a gas of strings to a dilaton-gravity background. In writing down the above action we have assumed that the string coupling is small, $g^2=e^{2\phi}\ll1$, that the curvature scalar $R_{10}\alpha^{\prime}\ll1$ and that field moduli evolve slowly (adiabaticity). There are branches of solutions for which these conditions are attractors as we shall demonstrate when the matter contribution, that is quantified by the equation of state parameter, $w$, varies slowly with size (see also \cite{Tseytlin:1991xk}). The quantity that we cannot neglect is the dilaton because this is required for the T-duality invariance of the theory. The equations of motion then follow from varying $S=S[G_{AB},\phi]$:
\begin{equation}\label{eq: eom1}
R^{A}_{\phantom{A}B} + 2\nabla^{A}\nabla_{B}\phi= \kappa_{10}^2e^{2\phi}T^{A}_{\phantom{A}B},
\end{equation}\\[-1cm]
\begin{equation}\label{eq: eom2}
R - 4(\nabla\phi)^2 + 4\nabla^2\phi= 0.
\end{equation}
We shall henceforth work in units where $\kappa_{10}^2=1/2$. Let us concentrate on homogeneous evolution of metric and on homogeneous and isotropic evolution of dilaton 
\begin{equation}
ds^2 = -dt^2+\sum_{i=1}^9e^{2\lambda_i(t)}dx_i^2,\quad \phi=\phi(t).
\end{equation}
The energy-momentum tensor is \emph{effectively}, as we shall demonstrate in section \ref{ST}, that of an ideal fluid but with variable equation of state,  $T^{A}_{\phantom{A}B}={\rm diag}(-\rho,p_1,\cdots,p_9)$. The equations of motion (\ref{eq: eom1}, \ref{eq: eom2}) then reduce to \cite{Tseytlin:1991xk} (see also \cite{Veneziano:1991ek}):
\begin{equation}\label{eq: constraint}
-\sum_{i=1}^9\dot{\lambda}_i^2+\dot{\varphi}^2=e^{\varphi+\sum_j\lambda_j}\rho,
\end{equation}\\[-1cm]
\begin{equation}\label{eq: ddot lambda}
\ddot{\lambda_i}-\dot{\varphi}\dot{\lambda_i}=\frac{1}{2}\,e^{\varphi+\sum_j\lambda_j}p_i,
\end{equation}\\[-1cm]
\begin{equation}\label{eq: ddot varphi}
\ddot{\varphi}-\sum_{i=1}^{9}\dot{\lambda_i}^2 = \frac{1}{2}\,e^{\varphi+\sum_j\lambda_j}\rho,
\end{equation}
where $\varphi$ is the re-scaled dilaton
\begin{equation}\label{eq: varphi}
\varphi\equiv2\phi-\sum_{i=1}^9\lambda_i.
\end{equation}
We are neglecting the effect of interactions and so we are studying the evolution of the universe after the hierarchy of scales has been created due to the BV mechanism. We are interested therefore in the case when $d$ dimensions are large, equal and free to expand and are assuming $9-d$ dimensions are static and at string scale (the case $d=3$ being of interest for the BV scenario). To see that this factorisation is consistent with the equations of motion note that at the string scale the pressure is negligibly small \cite{Brandenberger:1988aj,Tseytlin:1991xk,Bassett:2003ck}. Then, writing
\begin{equation*}
\lambda_i(t)=\left\{
\begin{array}{ll}
\lambda(t) & \textrm{for $i=\{1,\dots,d\}$,}\\
\mu(t) & \textrm{for $i=\{d+1,\dots,9\}$,}
\end{array} \right.
\end{equation*}
it follows from (\ref{eq: ddot lambda}) that $\mu(t)={\rm const}$ is a particular solution when $p_i|_{i\geq d+1}=0$. However, the pressure, $p_i$, is vanishing only at string scale and so it is consistent to take 
\begin{equation}\label{eq: lambda factorise}
\lambda_i(t)=\left\{
\begin{array}{ll}
\lambda(t) & \textrm{for $i=\{1,\dots,d\}$,}\\
0 & \textrm{for $i=\{d+1,\dots,9\}$.}
\end{array} \right.
\end{equation}
In making this deduction we have assumed that the equations of motion presented above hold at string scale and so this may not be a good assumption because non-perturbative effects are expected to become important at the string scale. The density and pressure of the large dimensions relate to the enery via $\rho(\lambda)=E/V$ and $p(\lambda)=-dE/dV$ respectively with $V=e^{d\lambda}$. Only two of the equations (\ref{eq: constraint}-\ref{eq: ddot varphi}) are independent ((\ref{eq: constraint}) being a constraint). These equations have been studied \cite{Tseytlin:1991xk,Bassett:2003ck} in some detail for $p\simeq0$ (string scale Hagedorn region) and $p=\rho/d$ (massless momentum mode or radiation domination) and so we shall concentrate on large radius Hagedorn regime where the pressure, $p$, is approximately constant and (as we shall show in section \ref{SEF2}) the variation of the equation of state cannot be neglected.

In the next section we recast the equations of motion into a form that makes them easier to solve analytically and determine their underlying structure.   

%%%%%%%%%%%%%%%%%%%%%%%%%%%%%%%%%%%%%%%%%%%%%%%%%%%%%%%%%%%%%%%%%%%%%%%%%

\section{Equations of Motion in String and Einstein Frame}\label{SEF1}
\subsection{String Frame}
\begin{figure}
\begin{center}
% This file is generated by the MATLAB m-file laprint.m. It can be included
% into LaTeX documents using the packages graphicx, color and psfrag.
% It is accompanied by a postscript file. A sample LaTeX file is:
%    \documentclass{article}\usepackage{graphicx,color,psfrag}
%    \begin{document}\input{dYdLp}\end{document}
% See http://www.mathworks.de/matlabcentral/fileexchange/loadFile.do?objectId=4638
% for recent versions of laprint.m.
%
% created by:           LaPrint version 3.16 (13.9.2004)
% created on:           14-Nov-2007 16:52:47
% eps bounding box:     9.375 cm x 7.0313 cm
% comment:              
%
\begin{psfrags}%
\psfragscanon%
%
% text strings:
\psfrag{s03}[t][t]{\color[rgb]{0,0,0}\setlength{\tabcolsep}{0pt}\begin{tabular}{c}$Y(\lambda)$\end{tabular}}%
\psfrag{s04}[b][b]{\color[rgb]{0,0,0}\setlength{\tabcolsep}{0pt}\begin{tabular}{c}${dY}/{d\lambda}$\end{tabular}}%
%
% xticklabels:
\psfrag{x01}[t][t]{$-1/w(\lambda)$}%
\psfrag{x02}[t][t]{$-d^{\phantom{0}^{\phantom{o}}}$}%
\psfrag{x03}[t][t]{$-\sqrt{d}$}%
\psfrag{x04}[t][t]{0}%
\psfrag{x05}[t][t]{$\sqrt{d}$}%
%
% yticklabels:
\psfrag{v01}[r][r]{0}%
%
% Figure:
\resizebox{7.5cm}{!}{\includegraphics{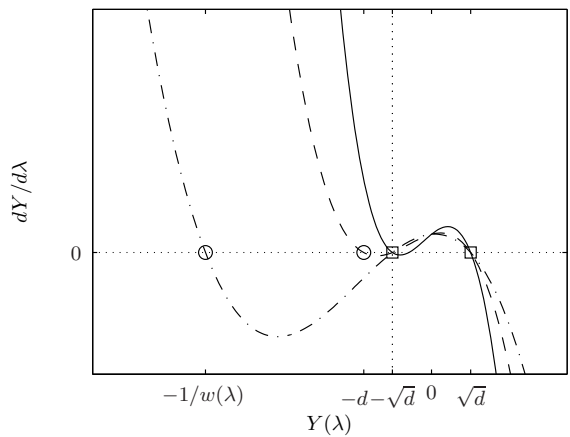}}%
\end{psfrags}%
%
% End dYdLp.tex

%\includegraphics[width=0.43\textwidth]{dYdL.eps}
\end{center}
\caption{\small{A flow diagram for $dY/d\lambda$. The dash-dot curve
represents the dependence of $dY/d\lambda$ on $Y$ for some value $0<w(\lambda)<1/d$,
while the dashed line represents this relation for $w=1/d$. This corresponds to the
asymptotic form of $Y^{\prime}$ vs $Y$ in an expanding universe
where $Y\leq-d$, see (\ref{eq: rho>0 condition},\ref{eq: dot-phi<0
condition}). The solid curve corresponds to the maximum value of $w$ allowed by the dominant energy condition, $|w|<1$. At $Y=\pm\sqrt{d}$ (represented by the squares) the
value of $Y^{\prime}$ is independent of $w$. We see that the
solution $Y=-1/w$ (represented by a circle) is attractive for
all initial $Y<-\sqrt{d}$ (provided $w$ varies slowly with size) and tends to $-\infty$ close to the self dual point where $w=\rightarrow0$.}}\label{fig: Y' vs Y}
\end{figure}

\begin{figure}
\begin{center}
% This file is generated by the MATLAB m-file laprint.m. It can be included
% into LaTeX documents using the packages graphicx, color and psfrag.
% It is accompanied by a postscript file. A sample LaTeX file is:
%    \documentclass{article}\usepackage{graphicx,color,psfrag}
%    \begin{document}\input{dHdLp}\end{document}
% See http://www.mathworks.de/matlabcentral/fileexchange/loadFile.do?objectId=4638
% for recent versions of laprint.m.
%
% created by:           LaPrint version 3.16 (13.9.2004)
% created on:           14-Nov-2007 16:55:59
% eps bounding box:     9.375 cm x 7.0313 cm
% comment:              
%
\begin{psfrags}%
\psfragscanon%
%
% text strings:
\psfrag{s03}[t][t]{\color[rgb]{0,0,0}\setlength{\tabcolsep}{0pt}\begin{tabular}{c}$Y(\lambda)$\end{tabular}}%
\psfrag{s04}[b][b]{\color[rgb]{0,0,0}\setlength{\tabcolsep}{0pt}\begin{tabular}{c}${d}(\ln H)/{d\lambda}$\end{tabular}}%
%
% xticklabels:
\psfrag{x01}[t][t]{$-1/w(\lambda)$}%
\psfrag{x02}[t][t]{$-d^{\phantom{0}^{\phantom{o}}}$}%
\psfrag{x03}[t][t]{$-\sqrt{d}$}%
\psfrag{x04}[t][t]{0}%
\psfrag{x05}[t][t]{$\sqrt{d}$}%
%
% yticklabels:
\psfrag{v01}[r][r]{}%
\psfrag{v02}[r][r]{0}%
\psfrag{v03}[r][r]{}%
%
% Figure:
\resizebox{7.5cm}{!}{\includegraphics{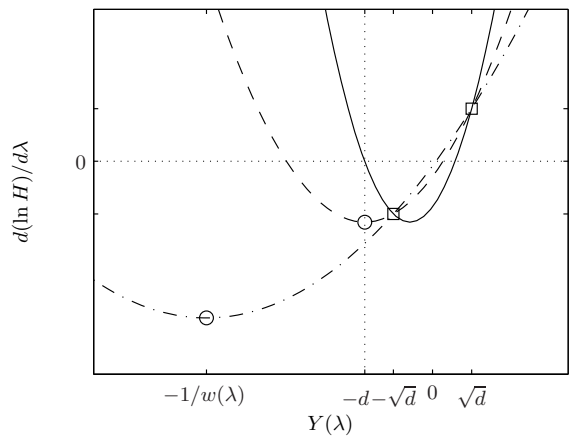}}%
\end{psfrags}%
%
% End dHdLp.tex

%\includegraphics[width=0.43\textwidth]{dHdL.eps}
\end{center}
\caption{\small{A sketch of $(\ln H)^{\prime}$ vs $Y$ (equation
(\ref{eq: Hubble exact})). The dash-dot curve represents this relation
for some value $0<w(\lambda)<1/d$, while the dashed line
represents this relation for $w=1/d$ (the asymptotic form of
$(\ln H)^{\prime}$ vs $Y$ in an expanding universe where $Y\leq-d$ (see
(\ref{eq: rho>0 condition},\ref{eq: dot-phi<0 condition}))).  The solid curve corresponds to the maximum value of $w$ allowed by the dominant energy condition, $|w|<1$. At
$Y=\pm\sqrt{d}$ (represented by the squares) the value of
$(\ln H)^{\prime}$ is independent of $w$. From Fig~\ref{fig: Y'
vs Y} we know that the solution $Y=-1/w$ (represented by a
circle at the minimum of the solid curve) is attractive for all
initial $Y<-\sqrt{d}$ provided $w$ varies slowly with size.  This minimum propagates from
$Y\rightarrow-\infty$, when $w\simeq0$, through to
$Y\rightarrow-d$, when $w=1/d$, in which case the dashed
and solid curves coincide. Whether $w>1/d$ is possible is not clear at present but the leading terms of the density of states seem to imply that $w>1/d$ \emph{is} possible. Note that when the density is below the Hagedorn density radiation is expected to dominate the spectrum and so $w=1/d$ eventually.}}\label{fig: H'/H vs Y}
\end{figure}

Let us identify $w(\lambda)$ with the equation of state parameter $p\equiv w(\lambda)\rho$. In section \ref{ST} we shall derive the form of $w(\lambda)$ from the microcanonical ensemble. This definition enables us to reduce the equations of motion to first order uncoupled form by introducing the two following variables 
\begin{equation}\label{eq: YH defn}
Y(\lambda)\equiv d\varphi/d\lambda\quad {\rm and} \quad H(\lambda)\equiv\dot{\lambda},
\end{equation}
where $H$ is evidently the (string frame) Hubble parameter. With these definitions the equations (\ref{eq: constraint}-\ref{eq: ddot varphi}) on account of (\ref{eq: lambda factorise}) lead to the following two independent equations of motion
\begin{equation}\label{eq: Y prime}
\frac{dY}{d\lambda} =
-\frac{1}{2}\left(1+wY\right)\left(Y^2-d\right),
\end{equation}\\[-0.8cm]
\begin{equation}\label{eq: Hubble exact}
\frac{1}{H}\frac{dH}{d\lambda}=Y+\frac{w}{2}\left(Y^2-d\right).
\end{equation}
The constraint (\ref{eq: constraint}) will serve to set the initial conditions and takes the form
\begin{equation}\label{eq: constr}
H^2=\frac{e^{2\phi}}{Y^2-d}\rho.
\end{equation}
It thus becomes clear that the dilaton, its rate of change and the number of dimensions determine the evolving gravitational coupling. The structure of (\ref{eq: Y prime}) is pictorially depicted in Fig~\ref{fig: Y' vs Y} while that of (\ref{eq: Hubble exact}) is shown in Fig~\ref{fig: H'/H vs Y}.

The evolution of the dilaton follows immediately from (\ref{eq: varphi}) and (\ref{eq: YH defn}):
\begin{equation}\label{eq: dilaton rate}
\frac{d\phi}{d\lambda} \equiv\frac{1}{2}(Y+d).
\end{equation}
To determine the constraints on $Y$ note that from the constraint equation we learn that positivity of the energy implies
\begin{equation}\label{eq: rho>0 condition}
\rho>0\,\Leftrightarrow Y^2>d.
\end{equation}
The equations of motion tell us that this condition is attractive. Furthermore, in order for the action (\ref{eq: action full}) to be justified we require the branch of solutions that does not allow runaway increasing dilaton in an expanding universe: $Y<-\sqrt{d}$. In fact the decreasing dilaton branch of solutions satisfies,
\begin{equation}\label{eq: dot-phi<0 condition}
\dot{\phi}<0\Leftrightarrow \left\{
\begin{array}{ll}
Y<-d & \textrm{if $H>0$,}\\
Y>-d & \textrm{if $H<0$,}
\end{array} \right.
\end{equation}
which follows from the definitions (\ref{eq: YH defn}) of $H$ and $Y$. The condition for an increasing dilaton, $\dot{\phi}>0$, can be reached by reversing the inequalities involving $H$. Notice that an \emph{increasing} dilaton is possible in an expanding universe when $Y>-d$ but when $Y<-\sqrt{d}$ it reaches a maximum and then decreases as the $Y=-d$ line is crossed.

The equations are invariant under T-duality: $\lambda\mapsto-\lambda$, as $T:Y\mapsto-Y$ and
$T:w\mapsto-w$. Hence, we shall consider only the $w>0$ branch as the dual branch $w<0$ can be reached with a T-duality transformation. We also conclude from (\ref{eq: Y prime}) that 
\begin{equation}\label{eq: Yad}
Y_{\rm ad}(\lambda)=-\frac{1}{w(\lambda)},
\end{equation}
is an attractive solution when $w(\lambda)$ varies sufficiently slowly with increasing $\lambda$. We shall refer to (\ref{eq: Yad}) as the \emph{adiabatic} solution. By sufficiently slowly we mean that $w^{-2}dw/d\lambda\ll1$ should be satisfied. In section \ref{SEF2} we show that this condition is not satisfied in the large radius Hagedorn regime (in fact $w^{-2}dw/d\lambda\sim d/w\gg1$) and so the adiabatic approximation is \emph{not} a good approximation. It will prove useful however in that it enables us to obtain analytical solutions by expanding $Y$ about $-1/w$. See also (\ref{eq: dot-phi<0 condition}) and the dash-dot line in the flow diagram Fig \ref{fig: Y' vs Y} from which the attractor solution can be read off. Notice also that this solution corresponds to the minimum of the dash-dot line in Fig \ref{fig: H'/H vs Y} where it is seen that close to the adiabatic solution the Hubble parameter decreases in an expanding universe. 

At radiation domination $\phi$ approaches a constant (as for radiation $w=1/d$, see also (\ref{eq: dilaton rate})) \footnote{Derivatives with respect to
$\lambda$ will occasionally be denoted by a prime: $^{\prime}\equiv
d/d\lambda$.}, $\phi^{\prime}\rightarrow0$, and hence
$Y\rightarrow-d$ (the dash-dash line in Fig~\ref{fig: Y' vs Y},\ref{fig: H'/H vs Y}).
 
\subsection{Einstein Frame}
The string frame metric is conformally related to related to the Einstein metric via 
\begin{equation}\label{eq: G_E vs G}
\tilde{G}_{\mu\nu} = {\rm exp}\left\{-\frac{4\phi}{d-1}\right\}G_{\mu\nu},
\end{equation}
as can be verified by direct substitution of the new metric into the action (\ref{eq: action full}). We shall throughout denote Einstein frame quantities with a tilde. It follows that the Einstein and string frame Hubble parameters are related by
\begin{equation}\label{eq: H_E vs H}
\tilde{H} = e^{\frac{2\phi}{d-1}}\left(\frac{-Y-1}{d-1}\right)H,
\end{equation}
from which, on account of (\ref{eq: Y prime},\ref{eq: Hubble exact}), the rate of change of $\tilde{H}$ with respect to $\tilde{\lambda}$ follows 
\begin{equation}\label{eq: Einstein Hubble}
\frac{1}{\tilde{H}}\frac{d\tilde{H}}{d\tilde{\lambda}} = -d + \frac{1}{2}(d-1)(1-w(\lambda))\frac{Y^2-d}{(Y+1)^2},
\end{equation}
where $\tilde{\lambda}=\lambda-2\phi/(d-1)$.
Notice that we have written (\ref{eq: Einstein Hubble}) in terms of string frame quantities $Y(\lambda),w(\lambda)$. We shall mainly concentrate on string frame calculations and we have written down (\ref{eq: G_E vs G},\ref{eq: H_E vs H},\ref{eq: Einstein Hubble}) to gain an idea of how the corresponding Einstein frame quantities evolve.
From (\ref{eq: H_E vs H}) and (\ref{eq: dot-phi<0 condition}) it follows that in the decreasing dilaton branch of solutions expansion in the string frame necessitates expansion in the Einstein frame. 

In the next section we derive the thermodynamic observables of interest starting from the microcanonical density of states.
We concentrate on large compact spaces and derive the dependence of $w(x)$ on $x\,(\simeq R^d/S$, the inverse entropy density) (\ref{eq: alpha(lambda)}) that will enable us to solve the equations of motion in the regions of interest. 

%%%%%%%%%%%%%%%%%%%%%%%%%%%%%%%%%%%%%%%%%%%%%%%%%%%%%%%%%%%%%
\section{Superstring Thermodynamics}\label{ST}
Thermodynamics of non-interacting superstring gases was studied extensively in the late 1980's but the work that is most relevant for the present paper is found in \cite{Brandenberger:1988aj,Deo:1989bv,Deo:1988jj,Deo:1991mp,Deo:1991af,O'Brien:1987pn,Bowick:1985az,Bowick:1989us,Tan:1988pt,Jain:1997ga} and more recently in \cite{Abel:1999rq,Brower:2002zx}. Focus has mainly been on the Hagedorn region for both small and large spaces, compact and non-compact topologies. Here we merely present the results for the case of a large toroidal background in the vicinity of the Hagedorn temperature, where the string density is higher than 1 in string units.

The microcanonical approach starts from the definition of the microcanonical density of states $\Omega(E)\equiv \sum_\alpha \delta(E-E_\alpha)$, with the sum taken over all states $\alpha$ of total energy $E_\alpha$. We shall concentrate on the case where all 9 spatial dimensions are toroidally compactified with $9-d$ dimensions at string scale and the remaining $d$ dimensions of equal size $R\gg1$.  

The leading two terms of the total density of states for a large and compact stringy universe when $E\gg\rho_1 R^d$ turns out \cite{Deo:1991mp} to be given by,
\begin{multline}\label{eq: Omega}
\Omega(E,R) = \beta_0 e^{\beta_0E + a_0 R^d + b_0R^{d-1}+\dots}\\
\times\left[1-\frac{[(\beta_0-\beta_1)E]^{2d-1}}{(2d-1)!}e^{-(\beta_0-\beta_1)(E-\rho_1R^d)}+\dots\right],
\end{multline}
where $R\gg1$ and $d>2$.  The constants $a_0, b_0$ and $\rho_1$ are $\mathcal{O}(1)$ and $\rho_1$ is identified with Hagedorn energy density. {The quantities $\beta_n$ (the cases $n=0,1$ being of immediate interest) are defined to be the inverse temperatures at which the analytically continued (single or multi-string) partition function is singular. In particular, the leading singularities for $d$ large equal radii were found to be located at \cite{Deo:1989bv,Deo:1988jj}
\begin{equation}\label{eq: beta_n}
\beta_n = \left(2\pi^2\alphap\right)^{1/2}\times\left[\left(1-\frac{n}{2\bar{R}^2}\right)^{1/2}+\\
\left(2-\frac{n}{2\bar{R}^2}\right)^{1/2}\right],
\end{equation}
for positive or vanishing integer $n$ subject to $n/2\bar{R}^2<1$ and  $\bar{R}\equiv R/\sqrt{\alpha^{\prime}}$ (note that $\beta_0>\beta_{n\neq0}$). The inverse temperature $\beta=\beta_0$ is by definition the inverse Hagedorn temperature. Therefore, $\beta_0-\beta_1\simeq\beta_0/(4\sqrt{2}R^2)$ when $1/2R^2\ll1$. Note that the second term in the brackets of (\ref{eq: Omega}) differs from \cite{Deo:1991mp}, and a more detailed derivation is given in \cite{DM07}. (The right hand side of equation (3.15) in \cite{Deo:1991mp} should have an additional factor of $[(\beta_0-\beta_1)/\beta_0]^{2d-1}\sim1/R^{4d-2}$.)}

The leading behaviour $\Omega\sim e^{\beta_0E}$ is well known in string theory and is due to the exponentially rising degeneracy of states associated with high energy oscillator modes of string (see e.g. \cite{Green:1987sp}). An important difference between (\ref{eq: Omega}) and the corresponding expression for all radii at string scale is the factor $a_0R^d$ that appears in the exponent. This is absent when all dimensions are of string scale and is due to the momentum modes of string \cite{Jain:1997ga}, the corresponding winding modes are subleading and their contribution can be understood by a T-duality transformation on $\Omega$. In particular, the corresponding expression for the density of states for an ideal gas of light point particles \cite{Deo:1991af} is $\Omega\sim\exp(\rho^{d/(d+1)}R^d)$. The proportionality factor in the case of particles is the density, $\rho$, and this varies with size. The reason as to why the coefficient of $R^d$ is constant in the string calculation (\ref{eq: Omega}) is that the momentum modes density is constant above Hagedorn energy densities (see also \cite{Sakellariadou:1995vk} for a classical simulation that supports this view). It is hence clear that the term $e^{a_0R^d}$ appears due to the presence of momentum modes, roughly speaking the centre of mass motion of short loops of string.

%%%%%%%%%%%%%%%%%%%%%%%%%%%%%%%%%%%%%%%%%%%%%%%%%%%%%%%%%%%%%%
\subsection{Equation of state}

From the entropy $S=\ln \Omega$ (up to an overall constant), all thermodynamic observables follow. 
From the assumption of adiabaticity, $S=\ln \Omega(E,R)={\rm const}$, with $\Omega(E,R)$ given by (\ref{eq: Omega}), we find that total energy of the string gas is 
\begin{equation}\label{eq: E}
E(x)= \beta_0^{-1}S(1-x-\mathcal{O}(x/R)), 
\end{equation}
where we have found it convenient to define $x\equiv a_0R^d/S$, which is essentially the inverse entropy density. This expression is valid for $R\gg1$ and $x\ll1$.

The pressure of the string gas system is given by $p=\beta^{-1}(\partial S/\partial V)_E$, with $V=R^d$. In particular, for the large volume string gas we find that the pressure
\begin{equation}\label{eq: p(x)}
p(x)= a_0\beta_0^{-1}+\frac{d-1}{d}\frac{b_0}{R}-\frac{2\rho}{d}\frac{\beta_0-\beta_1}{\beta_0}\delta\Omega+\dots,
\end{equation}
where $$\delta\Omega\equiv-\frac{[(\beta_0-\beta_1)E]^{2d-1}}{(2d-1)!}e^{-(\beta_0-\beta_1)(E-\rho_1R^d)}.$$
This result strictly applies in the region $\rho\gg\rho_1$. {Note however that the second term in the expression for the pressure is a correction that arises due to a higher order term in the leading exponential of (\ref{eq: Omega}) while the third term is associated to $\delta\Omega$.  In particular, there are in principle other corrections that are subleading relative to the second term but \emph{larger} than the third term which also come from the leading exponential in (\ref{eq: Omega}). The reason we have written the expression for $p(x)$ in the above form is because the corrections from the higher order terms in the exponential decrease with increasing radius whereas $\delta\Omega$ increases.  Therefore, a careful treatment of the subleading contributions is needed, especially close to the Hagedorn density, and we hope to pursue this in our follow-up paper \cite{DM07}. For the current paper the important term is the leading contribution while the sub-leading corrections have only been written down to obtain an order of magnitude estimate of the region of validity of the leading term.}
It is important to emphasise that the pressure does not vanish in this large radius Hagedorn regime (even though $p/\rho\sim x$) but rather is of order $a_0\beta_0^{-1}$. We find that the equation of state parameter, $w\equiv p/\rho$, is correspondingly given by
\begin{equation}\label{eq: alpha(lambda)}
w(x) = \frac{x}{1-x}\left[1+(bR^{-1})\frac{x}{1-x}\right]-\frac{2}{d}\frac{\beta_0-\beta_1}{\beta_0}\delta\Omega+\dots,
\end{equation}
where $b$ is an $\mathcal{O}(1)$ constant. This expression holds for a non-interacting string gas and breaks down at a critical volume $V_c\sim S$ (or $x_c\sim1$) in which case $\delta\Omega\sim1$. In what follows we shall only consider the  leading term. 

{For very large volumes the universe will be radiation dominated and $w\simeq1/d$. The equation of state in the Hagedorn/radiation transition region $x\sim1$ has not been determined to date, the calculation of which boils down to determining a sum of contour integrals of the analytically continued multi-string partition function around each of the singularities $\beta_n$. Some progress is made in this direction in \cite{DM07}. Furthermore, we have mentioned that the pressure close to the string scale, $R\sim1$ (in string units), vanishes for the string gas whereas we have shown in (\ref{eq: p(x)}) that for large radii it is of order 1. The dependence of the string gas pressure on size between these two extreme cases is also puzzling and has not been determined to date. However, we would rather not say much about the small radius region where the validity of the free string approach is questionable. In particular, one would expect strong coupling effects to become important.}

%%%%%%%%%%%%%%%%%%%%%%%%%%%%%%%%%%%%%%%%%%%%%%%%%%%%%%%%%%%
\subsection{String number density}
One can also estimate the number density of small loops and long winding strings. The total number of strings in an ensemble with individual energies in the range $\epsilon\rightarrow\epsilon+d\epsilon$ is classically given by \cite{Deo:1989bv}
\begin{equation}\label{eq: D(e;E)}
\mathcal{D}(\epsilon;E)\,d\epsilon=\frac{1}{\Omega(E)}\,f(\epsilon)\,\Omega(E-\epsilon)\,d\epsilon.
\end{equation}
$f(\epsilon)$ is the single string density of states, $f(\epsilon)\equiv\sum_\alpha\delta(\epsilon-\epsilon_\alpha)$ and $\epsilon_\alpha$ is the energy of the single string state $\alpha$.  In \cite{Deo:1991mp} it is shown that (for single string energies greater than some lower cut-off $\epsilon_0\sim{\alpha^{\prime}}^{-1/2}$)
\begin{equation}\label{eq: f}
f(\epsilon) = \sum_n g_n\frac{e^{\beta_n\epsilon}}{\epsilon},
\end{equation}%%%%%%%%%%%%%
{where the multiplicity of states, $g_n$, is defined to be the total number of possible configurations for which $n=\sum_{i=1}^dm_i^2$ for all $m_i\in \mathbb{Z}$. For example, $g_0=1$, $g_1=2d$, $g_2=2^2d(d-1)/2!$, $g_3=2^3d(d-1)(d-2)/3!$ and so on, whereas for very large $n$ we can make a continuum approximation and find 
\begin{equation}\label{eq: g_n}
g_n\simeq \frac{2\pi^{d/2}}{\Gamma(d/2)}\,n^{d/2}. 
\end{equation}
This corresponds to the number of states contained in a spherical shell of thickness $1$ and radius  $n^{1/2}$ \footnote{We have defined the radius to be $\sqrt{n}$ because we want the index $n$ to run over all integers. This simplifies things greatly when examining higher order poles in which case the index $n$ on $\beta_n$ takes only integer values.}. For the rest of this section we find it convenient to set $\alpha^{\prime}=1$.}%%%%%%%%%%

Equation (\ref{eq: f}) can then be evaluated for large and small single string energies leading to \cite{Deo:1991mp}
\begin{eqnarray}
f(\epsilon)&\simeq& R^d\frac{e^{\beta_0\epsilon}}{\epsilon^{d/2+1}},\quad \epsilon_0<\epsilon\ll R^2\label{eq: f(e1)}\\
&\simeq& \frac{e^{\beta_0\epsilon}}{\epsilon},\quad \epsilon\gg R^2\label{eq: f(e2)},
\end{eqnarray}
where $\epsilon_0$ is some cut-off ($\sim1$) below which our expression for density of states breaks down as mentioned above. {The first of these is calculated using the continuum approximation for the density of states, $g_n$, (\ref{eq: g_n}) where $n$ is large. This approximation can be made because, as can be seen from (\ref{eq: beta_n}), the radius-dependent singularities $\beta_{n>0}$ all approach $\beta_0$ (from below) as the radius $R$ increases \cite{Deo:1988jj} thus making the monotonic map $\beta_n:\mathbb{Z}\rightarrow \mathbb{R}$ approximately continuous and slowly varying. Therefore, all factors of the form $e^{{\beta_{n>0}}\epsilon}/\epsilon$ will be comparable for arbitrary $n$ provided $R^2\gtrsim 2n$, and so given that $g_n$ increases with increasing $n$ we expect the large $n$ region of $g_n$ to dominate the sum in (\ref{eq: f}) where the continuum approximation (\ref{eq: g_n}) can be made. The sum in (\ref{eq: f}) is therefore to be  replaced by an integral which can be performed explicitly to yield (\ref{eq: f(e1)}). The second of these (\ref{eq: f(e2)}) is calculated in the small radius limit. Here the leading (Hagedorn) singularity is far from any sub-leading singularities $\beta_{n>0}$ so that terms such as $e^{{\beta_{n>0}}\epsilon}/\epsilon$ are exponentially suppressed compared to $e^{\beta_{0}\epsilon}/\epsilon$. Hence, only the leading ($n=0$) term of the sum in (\ref{eq: f}) remains for large single string energies (\ref{eq: f(e2)}).} %%%%%%%%%%%

The high energy strings with $\epsilon\gg R^2$ correspond to long winding strings which can accommodate a larger number of oscillators compared to small loops which in turn correspond to low energies $\epsilon\ll R^2$. Note furthermore that one can interpret states with $\epsilon\sim R^2$ as random walks of step length $1$, whose size is of order $R$. 

{We can estimate the total number of strings, $N(E)$, in the ensemble from (\ref{eq: D(e;E)}) 
\begin{eqnarray}\label{eq: number}
N(E)&=&\frac{1}{\Omega(E)}\int_{0}^Ed\epsilon f(\epsilon)\Omega(E-\epsilon)\nonumber\\
&\simeq&\int_{\epsilon_0}^Ed\epsilon f(\epsilon)e^{-\beta_0\epsilon}.
\end{eqnarray}
Let us then write this integral as a sum of two contributions on account of (\ref{eq: f(e1)}) and  (\ref{eq: f(e2)}) to find that
\begin{eqnarray}\label{eq: tns}
N(E)&\sim& R^d\int_{\epsilon_0}^{R^2}\frac{d\epsilon}{\epsilon^{d/2+1}}+\int_{R^2}^E \frac{d\epsilon}{\epsilon}\nonumber\\
&\sim& R^d+\ln (E/R^2)+\dots
\end{eqnarray}
Therefore, from (\ref{eq: tns}) and (\ref{eq: E}) we find that }%%%%%%%%%%%% 
the number density of strings for a space with $d$ large compact dimensions will be
\begin{equation}\label{eq: nds}
n(R)\sim 1+R^{-d}\ln S.
\end{equation}
Roughly speaking, the first term counts the number density of small loops (coming from the $\epsilon\ll R^2$ term) while the second term counts the long string contribution (coming from the $\epsilon\gg R^2$ term) (see \cite{Deo:1991mp}). There is hence about one loop per unit volume and we see that long strings are energetically favourable at radii smaller than $\sim(\ln S)^{1/d}$. For radii larger than $\sim(\ln S)^{1/d}$ small loops will dominate the number density and the contribution from long string is expected to be small. Furthermore, at a critical radius $R_r\sim S^{1/d}$ there is not enough energy to excite the massive states however and massless momentum modes (radiation) will come to dominate \cite{Brandenberger:1988aj} the ensemble. This radius, $R_r$, is reached when the total energy density is of order one, or $E\sim\rho_1 R^d$, and it is seen from (\ref{eq: Omega}) that this is also where our calculations break down; in fact higher order terms in the expansion of $\Omega$ also become important at this density as one would expect \cite{DM07}. {Notice that the number density of windings, $n_w$, will be completely negligible at this critical radius $R_r\sim S^{1/d}$: 
\begin{equation}\label{eq: n_w^c}
n_w^{c}\sim (\ln S)/S\ll1. 
\end{equation}
}

{We have thus found that there are three distinct regions in the large radius regime. For radii $R$ in the range
\begin{equation*}
1\ll R\ll(\ln S)^{1/d},
\end{equation*}
there is a large number of long winding string compared to small loops. For radii in the range
\begin{equation*}
(\ln S)^{1/d}\ll R\ll S^{1/d},
\end{equation*}
the large radius Hagedorn region that is most relevant for the rest of this paper,
there is expected to be only a small number of long string compared to small loops which dominate the number density. Note that this transition from long to short string does not affect the equations of state. Finally, for radii $R$ in the range
\begin{equation*}
S^{1/d}\ll R,
\end{equation*}
i.e. outside the large radius Hagedorn region, there is not enough energy to excite massive modes and the spectrum is dominated by the ground state which is comprised of massless modes, i.e. radiation. This latter transition does affect the equation of state.}

%%%%%%%%%%%%%%%%%%%%%%%%%%%%%%%%%%%%%%%%%%%%%%%%%%%%%%%%%%%%%%%%%%%%%%%%%
\section{Evolution in String Frame}\label{SEF2}
\parindent=2 em 
In this section we shall present solutions for the Hubble parameter and the dilaton as a function of size for the equation of state parameter (\ref{eq: alpha(lambda)}) that was derived in the previous section. Keeping only the leading contribution we have
\begin{equation}\label{eq: w}
w(x) \simeq \frac{x}{1-x}+\dots
\end{equation}
Recall that $x\equiv a_0 R^d/S\ll1$ is the inverse entropy density multiplied by an $\mathcal{O}(1)$ constant $a_0>0$.

As a first approximation let us consider the adiabatic attractor solution (\ref{eq: Yad}) (in an expanding universe with decreasing dilaton)
\begin{displaymath}
Y_{\rm ad}(x)=-\frac{1}{w_{\rm ad}(x)}.
\end{displaymath}
It becomes clear that we cannot identify $w_{\rm ad}(x)$ with (\ref{eq: w}) because the condition that we trust the adiabatic approximation, $w^{-2}|dw/d\lambda|\ll1$, is violated for the equation of state parameter (\ref{eq: w}), for which $w^{-2}dw/d\lambda\sim d/w\gg1$. Furthermore, this is true for the entire range of $w$ allowed by the dominant energy condition. Let us then expand about the adiabatic approximation in the following manner:
\begin{equation}\label{eq: Y_ep}
Y(x)\equiv -\frac{1}{w(x)}(1+\epsilon(x)).
\end{equation}
On account of the equation of motion of $Y(\lambda)$ (\ref{eq: Y prime}), we obtain the equation of motion of $\epsilon(x)$. We shall find that $\epsilon$ increases with increasing $w$ and shall hence refer to the solution (\ref{eq: Y_ep}) as the \emph{non-adiabatic} approximation. We find it convenient to take as our independent variable the equation of state parameter $w$; on account of (\ref{eq: w}), $x=w/(1+w)$.  Then, the equation of motion for $\epsilon(w)$ is
\begin{equation}\label{eq: epsilon}
\frac{d\epsilon(w)}{dw} \simeq \frac{1+\epsilon}{w}\left[ 1-\frac{1}{2d}\frac{\epsilon(1+\epsilon)}{w(1+w)}\right]+\dots, 
\end{equation}
where $"\dots"$ denote terms of higher order in $w$. Note that $\epsilon(w)$ is not necessarily small and that $|w|<1$.  From (\ref{eq: Hubble exact}) we see that the differential equation for the string frame Hubble parameter $H$ can correspondingly be written in terms of $\epsilon(w)$ as follows
\begin{equation}\label{eq: dHubble_nonad}
\frac{d}{dw}\ln H=\frac{1}{2(1+w)}\left[\frac{\epsilon^2-1}{dw^2}-1\right].
\end{equation}
The first thing to notice is that there are two opposing terms in (\ref{eq: epsilon}).  If $\epsilon/w\ll 2d$ then there is a runaway solution so that the adiabatic solution $\epsilon=0$ is repulsive. On the other hand, when $\epsilon/w\gg2d$ the $\epsilon=0$ solution becomes attractive. \emph{Independently} of which of these two initial conditions the universe chooses the attractor of (\ref{eq: epsilon}) has approximately the following form
\begin{equation}\label{eq: epw_apx}
\epsilon(w)\simeq \frac{1}{2}\left[\sqrt{1+8dw}-1\right].
\end{equation}
Furthermore, it follows from (\ref{eq: dHubble_nonad}) that when $\epsilon\gtrsim1+\mathcal{O}(w^2)$ the Hubble parameter \emph{increases} with increasing size. This, on account of (\ref{eq: epw_apx}), will be the case when $w\gtrsim1/d$. In other words, if $w$ overshoots the momentum mode value, $1/d$, the string frame Hubble parameter will increase. 

\begin{figure}
\begin{center}
% This file is generated by the MATLAB m-file laprint.m. It can be included
% into LaTeX documents using the packages graphicx, color and psfrag.
% It is accompanied by a postscript file. A sample LaTeX file is:
%    \documentclass{article}\usepackage{graphicx,color,psfrag}
%    \begin{document}\input{phix}\end{document}
% See http://www.mathworks.de/matlabcentral/fileexchange/loadFile.do?objectId=4638
% for recent versions of laprint.m.
%
% created by:           LaPrint version 3.16 (13.9.2004)
% created on:           14-Nov-2007 16:06:35
% eps bounding box:     9.375 cm x 7.0313 cm
% comment:              
%
\begin{psfrags}%
\psfragscanon%
%
% text strings:
\psfrag{s03}[b][b]{\color[rgb]{0,0,0}\setlength{\tabcolsep}{0pt}\begin{tabular}{c}$\phi(x)-\phi(x_0)$\end{tabular}}%
\psfrag{s04}[t][t]{\color[rgb]{0,0,0}\setlength{\tabcolsep}{0pt}\begin{tabular}{c}$x$\end{tabular}}%
%
% xticklabels:
\psfrag{x01}[t][t]{0.1}%
\psfrag{x02}[t][t]{0.2}%
\psfrag{x03}[t][t]{0.3}%
\psfrag{x04}[t][t]{0.4}%
\psfrag{x05}[t][t]{0.5}%
%
% yticklabels:
\psfrag{v01}[r][r]{-1.4}%
\psfrag{v02}[r][r]{-1.2}%
\psfrag{v03}[r][r]{-1}%
\psfrag{v04}[r][r]{-0.8}%
\psfrag{v05}[r][r]{-0.6}%
\psfrag{v06}[r][r]{-0.4}%
\psfrag{v07}[r][r]{-0.2}%
\psfrag{v08}[r][r]{0}%
%
% Figure:
\resizebox{7.5cm}{!}{\includegraphics{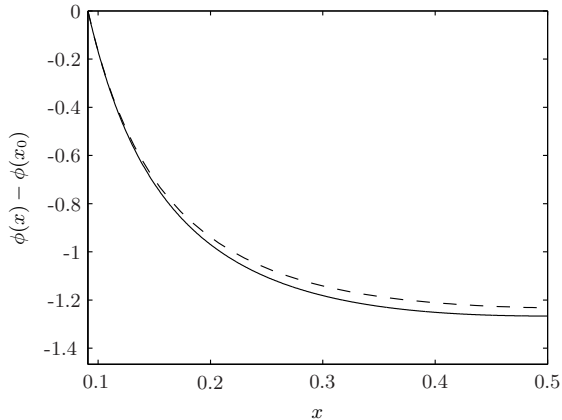}}%
\end{psfrags}%
%
% End phix.tex

%\includegraphics[width=0.43\textwidth]{phi.eps}
\end{center}
\caption{\small{A plot of the dilaton, $\phi(x)$, versus $x$ for $w_0=0.1$. The dash-dash line corresponds to the full numerical solution while the solid line corresponds to the analytic solution (\ref{eq: dil_nonad}).}}\label{fig: phix}
\end{figure}

\begin{figure}
\begin{center}
% This file is generated by the MATLAB m-file laprint.m. It can be included
% into LaTeX documents using the packages graphicx, color and psfrag.
% It is accompanied by a postscript file. A sample LaTeX file is:
%    \documentclass{article}\usepackage{graphicx,color,psfrag}
%    \begin{document}\input{Hx}\end{document}
% See http://www.mathworks.de/matlabcentral/fileexchange/loadFile.do?objectId=4638
% for recent versions of laprint.m.
%
% created by:           LaPrint version 3.16 (13.9.2004)
% created on:           14-Nov-2007 16:13:57
% eps bounding box:     9.375 cm x 7.0313 cm
% comment:              
%
\begin{psfrags}%
\psfragscanon%
%
% text strings:
\psfrag{s03}[b][b]{\color[rgb]{0,0,0}\setlength{\tabcolsep}{0pt}\begin{tabular}{c}$H(x)/H(x_0)$\end{tabular}}%
\psfrag{s04}[t][t]{\color[rgb]{0,0,0}\setlength{\tabcolsep}{0pt}\begin{tabular}{c}$x$\end{tabular}}%
%
% xticklabels:
\psfrag{x01}[t][t]{0.1}%
\psfrag{x02}[t][t]{0.2}%
\psfrag{x03}[t][t]{0.3}%
\psfrag{x04}[t][t]{0.4}%
\psfrag{x05}[t][t]{0.5}%
%
% yticklabels:
\psfrag{v01}[r][r]{0.4}%
\psfrag{v02}[r][r]{0.5}%
\psfrag{v03}[r][r]{0.6}%
\psfrag{v04}[r][r]{0.7}%
\psfrag{v05}[r][r]{0.8}%
\psfrag{v06}[r][r]{0.9}%
\psfrag{v07}[r][r]{1}%
%
% Figure:
\resizebox{7.5cm}{!}{\includegraphics{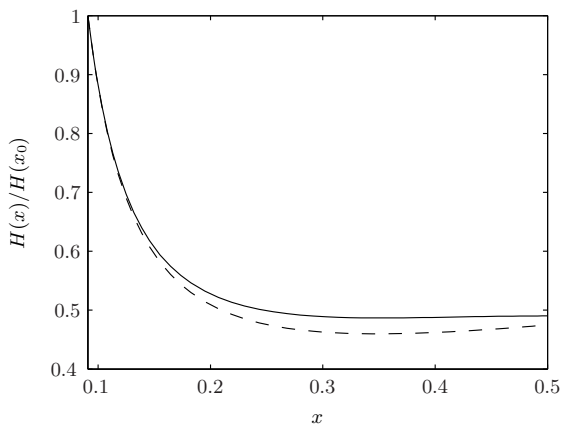}}%
\end{psfrags}%
%
% End Hx.tex

%\includegraphics[width=0.43\textwidth]{H.eps}
\end{center}
\caption{\small{A plot of the Hubble parameter, $H(x)$, versus $x$ for $w_0=0.1$. The dash-dash line corresponds to the numerical solution while the solid line corresponds to the analytic solution (\ref{eq: Hubble_nonad}). Notice that there is a minimum $H(x)$ when $w\simeq 1/d$ so that $H$ actually slightly increases beyond this value until $x\simeq1/2$ where $w\simeq 1$.}}\label{fig: Hw}
\end{figure}

We have shown that the true evolution departs from the adiabatic approximation, $Y\simeq-1/w$, by a factor of approximately $(\sqrt{1+8dw}+1)/2$. We can however derive the evolution of the dilaton from the definitions (\ref{eq: Y_ep}) and (\ref{eq: dilaton rate}) for the solution (\ref{eq: epw_apx}) and we find that
\begin{multline}\label{eq: dil_nonad}
\phi(w)\simeq \phi(w_0)+\frac{\sqrt{1+8dw}+1}{4dw}+\\
\ln\left[\left(\frac{\sqrt{1+8dw}+1}{\sqrt{1+8dw}-1}\right)^{1-\frac{1}{4d}}\left(\frac{w}{1+w}\right)^{\frac{1}{2}+\frac{1}{4d}}\right]+\\
\frac{\sqrt{8d-1}}{2d}\arctan\sqrt{\frac{1+8dw}{8d-1}}+f_1(w_0),
\end{multline}
where $f_1(w_0)$ is an integration constant such that $\phi(w)|_{w_0}=\phi(w_0)$. This equation tells us that the dilaton decreases rapidly and monotonically with increasing equation of state parameter $w$. In the limit $w\ll1$ the dilaton decreases according to 
\begin{equation}\label{eq: phi smallw}
\phi(w)-\phi(w_0)\simeq -\frac{1}{2d}\left(\frac{1}{w_0}-\frac{1}{w}\right)+\dots,
\end{equation}
where \texttt{"}$\dots$\texttt{"} denote higher order terms in $w$. The solution (\ref{eq: dil_nonad}) is plotted alongside the numerical solution (by which we mean the numerical solution to equation (\ref{eq: Y prime}) when sourced by (\ref{eq: w})) in Fig \ref{fig: phix}.
We can likewise determine the evolution of the string frame Hubble parameter and from (\ref{eq: epw_apx}), (\ref{eq: Y_ep}) and (\ref{eq: Hubble exact}) we find that 
\begin{multline}\label{eq: Hubble_nonad}
\ln\frac{H(w)}{H(w_0)}=\frac{\sqrt{1+8dw}+1}{4dw} \\
+\ln\left[\left(\frac{w}{1+w}\right)^{1+\frac{1}{4d}}\left(\frac{\sqrt{1+8dw}+1}{\sqrt{1+8dw}-1}\right)^{1-\frac{1}{4d}}(1+w)^{-1/2}\right]\\
+\frac{\sqrt{8d-1}}{2d}\arctan\sqrt{\frac{1+8dw}{8d-1}}+f_2(w_0),
\end{multline}
where $f_2(w_0)$ is again an integration constant analogous to $f_1(w_0)$. This solution is plotted in Fig~\ref{fig: Hw} alongside the numerical solution.
One then sees that to leading order in $w$,
\begin{equation}\label{eq: Hsolapx}
H(w)\simeq H_0\exp\left\{-\frac{1}{2d}\left[\frac{1}{w_0}-\frac{1}{w}\right]\right\}\left(\frac{w}{w_0}\right)^{\frac{1}{2d}}+\dots,
\end{equation}
where the \texttt{"}$\dots$\texttt{"} higher order terms in $w$. Hence, we have found that in the large radius Hagedorn regime the Hubble parameter \emph{decreases} at an ever slower rate until $w\sim1/d$, beyond which it slightly increases and this result is \emph{independent} of the initial conditions, i.e. independent of whether $(|\epsilon|/w)_{\rm initial}$ is smaller or greater than $\mathcal{O}(2d)$. We can further obtain the approximate time evolution of radius; it is more convenient to work in terms of entropy density, $w\simeq x=a_0/s$. After some algebraic manipulations (\ref{eq: Hsolapx}) can be integrated. The resulting incomplete gamma functions can be expanded about large $s,s_0$; keeping leading order terms only we find that
\begin{equation}\label{eq: xt}
\frac{R}{R_{\rm in}}\simeq \left\{1-\frac{2da_0}{s_{\rm in}}\ln \left(1+\frac{s_{\rm in}H_{\rm in}\Delta t}{2a_0}\right)\right\}^{-1/d}.
\end{equation}
The constant $a_0$ indicates that it is the momentum modes (as well as the oscillator modes) that play an important role in the evolution of the large radius universe rather than the winding modes. Roughly speaking, in the absence of momentum modes, $a_0=0$, and hence the universe remains static to this order. The initial entropy density is $s_{\rm in}\equiv S/R_{\rm in}^d$ and one can think of $R_{\rm in}$ (although not necessarily) as the radius of the universe beyond which there is a negligible number of long string (\ref{eq: nds}), $$R_{\rm in}\sim (\ln S)^{1/d}\gg1,$$ which can be thought of as the radius that distinguishes large from small radius Hagedorn phase.

The characteristic timescale of the Hagedorn phase can be estimated from (\ref{eq: xt}), $\Delta t_H\sim e^{s_{\rm in}}/(s_{\rm in}H_{\rm in})$. Note that we can further eliminate the dependence on the initial Hubble parameter on account of (\ref{eq: H0}) in favour of the initial string coupling $g_{\rm in}=e^{\phi_{\rm in}}$. Therefore, the duration of the Hagedorn phase is estimated to be
\begin{equation}\label{eq: t_H}
\Delta t_H\sim \frac{1}{g_{\rm in}}\frac{e^{s_{\rm in}}}{\sqrt{s_{\rm in}}}.
\end{equation}
As one would expect, in the limit of infinite entropy density \mbox{$s_{\rm in}\rightarrow\infty$} one finds that the universe is static $R(t)=R_{\rm in}$ while $\Delta t_H\rightarrow\infty$. 

The dilaton time evolution can also be computed and from (\ref{eq: phi smallw}) and (\ref{eq: xt}) it follows that
\begin{equation}\label{eq: phit}
\phi(t)-\phi(t_{\rm in})\simeq -\ln \left(1+\frac{s_{\rm in}H_{\rm in}\Delta t}{2a_0}\right)+\dots
\end{equation}
To see that the dilaton runs to very weak coupling notice that (\ref{eq: phit}) in terms of string coupling and for large times takes the form $g\sim g_{\rm in} (s_{\rm in}H_{\rm in}\Delta t)^{-1}$. Then, on account of (\ref{eq: t_H}) and (\ref{eq: H0}) we find that at the onset of radiation domination the string coupling, $g=e^{\phi}$, will have reached
\begin{equation}\label{eq: gr}
g_r\sim g_{\rm in}e^{-s_{\rm in}},
\end{equation}
which for reasonable values of the initial entropy density is tiny on account of $g_{\rm in}\lesssim 1$.

We shall not write down the dependence of the Einstein frame Hubble parameter on Einstein frame volume, this amounts to substituting (\ref{eq: Hubble_nonad},\ref{eq: dil_nonad},\ref{eq: epw_apx},\ref{eq: Y_ep}) into (\ref{eq: H_E vs H}). It suffices to say that the Einstein frame Hubble parameter, $\tilde{H}$, always decreases with increasing (Einstein frame) volume unless the dominant energy condition is violated, $w\leq-1$, (see (\ref{eq: Einstein Hubble})) or positivity of energy condition, $Y^2>d$, is relaxed. Furthermore, for $\tilde{H}$ to increase for a large range of values for $Y$ (rather than just close to $Y\sim -d$) the condition becomes more restrictive: $w<-(d+1)/(d-1)$. These results are in accordance with \cite{Kaloper:2006xw}. It is also important to note that positive energy and decreasing dilaton necessitate $H$ and $\tilde{H}$ to have the same sign, so that if the universe expands in the string frame it also expands in the Einstein frame. 

%%%%%%%%%%%%%%%%%%%%%%%%%%%%%%%%%%%%%%%%%%%%%%%%%%%%%%%%%%%%%%%%%%%%%%
\section{Problems of the String Gas}\label{SEP}
We shall now discuss various problems of the dilaton-gravity realisation of string gas cosmology. We shall first consider the issue of thermal equilibrium and improve on the current estimates found in \cite{Danos:2004jz} (and also \cite{Takamizu:2006sy}) by showing that the assumption of thermal equilibrium indeed is not justifiable. We provide sufficient tools to actually estimate the evolution of the interaction rate per Hubble rate as a function of size (or equivalently entropy density $s$) and present an analytic estimate of this quantity in the limit $w\ll1$. We then concentrate on the late time phenomenology of this scenario. By late time we mean radiation domination onwards. {In what follows we make the assumption that the end of the Hagedorn era (where the universe is dominated by massless modes of string) is to be identified with the radiation dominated phase of standard Big Bang cosmology in order to extract late time predictions from this model.}

In the current section we show that the Hubble length must shrink to a tiny fraction of the size of the universe for a toroidal background, on which the above scenario is based, to be acceptable. We then go on to show that there is a dilaton problem associated with this scenario, namely that post-Hagedorn considerations require the string coupling to be $\mathcal{O}(1)$ at the onset of radiation, while the Hagedorn evolution leads to a very weakly coupled dilaton. In addition to these problems there is also an entropy problem, which stems from the fact that in the absence of some additional mechanism (e.g. inflation) string gas cosmology does not offer an explanation for the large amount of entropy in the current day universe, although it has been speculated that the entropy may be generated by a period of oscillation around the self-dual radius \cite{Tseytlin:1991xk} (see also \cite{Kaloper:2006xw,Biswas:2005qr,Brandenberger:2005qj,Biswas:2006bs,Biswas:2008ti}).
%%%%%%%%%%%%%%%%%%%%%%%%%%%%%%%%%%%%%%%%%%%%%%%%%%%%%%%%%%%%%%

\subsection{Thermal Equilibrium and Jeans Instabilities}
The question of thermal equilibrium has been addressed in \cite{Danos:2004jz} (see also \cite{Takamizu:2006sy}) in the deep Hagedorn regime, namely at the self dual point.  An estimate was obtained and this indicated that thermal equilibrium cannot be maintained. However, at the self-dual point the universe may well be strongly coupled which leads us to distrust the dilaton-gravity action from which we started from.  In particular, it is only meaningful to speak about thermal equilibrium at epochs where we can trust our theory. We shall here only consider the large volume region where the formalism developed above applies. The quantity of interest is the ratio of interaction rate per expansion rate, $\Gamma/H$. For thermal equilibrium we require this ratio to be much greater than 1. We can estimate the interaction rate at large radii from
\begin{equation}\label{eq: Gamma}
\frac{\Gamma}{H}\sim\frac{n}{H}e^{4\phi},
\end{equation}
where $n$ is the number density of strings. This was found in section \ref{ST} to be given by
\begin{displaymath}
n(R)\sim 1+R^{-d}\ln S.
\end{displaymath}
We shall not display the full result but shall concentrate on the limiting case $w\ll1$. Then, with $w\simeq x=a_0/s$, with $s$ the entropy density, the Hubble parameter (\ref{eq: Hsolapx}) and dilaton (\ref{eq: phi smallw}) can be used to estimate the dependence of $\Gamma/H$ on $s(=w/a_0)$:
\begin{displaymath}
\frac{\Gamma}{H}\sim H_{\rm in}^{-1}e^{4\phi_{\rm in}}\left(\frac{s}{s_{\rm in}}\right)^{1/2d}\,e^{-3(s_{\rm in}-s)/2d},
\end{displaymath}
where we have taken $R\gtrsim(\ln S)^{1/d}$ so that $n(s)\sim 1$ as appropriate for the large radius solutions for $\phi$ and $H$. We can do better than this however because (\ref{eq: constr}) constrains the initial conditions as was emphasised in section~\ref{SEF1}, $H^2_{\rm in}\simeq e^{2\phi_{\rm in}}\rho_{\rm in}/Y_{\rm in}^2$. In particular, $w_{\rm in}\ll1$ and so (\ref{eq: Y_ep}) on account of (\ref{eq: epw_apx}) enforces $Y_{\rm in}=-1/w_{\rm in}\simeq-s_{\rm in}/a_0$. Furthermore, $\rho_{\rm in}=1/x_{\rm in}= s_{\rm in}/a_0$ and hence the initial Hubble parameter must satisfy 
\begin{equation}\label{eq: H0}
H_{\rm in}\simeq e^{\phi_{\rm in}}(a_0/s_{\rm in})^{1/2},
\end{equation}
from which we see that we can trade the initial value of the Hubble constant for the initial value of the dilaton. Substituting this into the above estimate implies that
\begin{equation}\label{eq: intrate}
\frac{\Gamma}{H}\sim e^{3\phi_{\rm in}}\left(\frac{s_{\rm in}}{a_0}\right)^{1/2}\left(\frac{s}{s_{\rm in}}\right)^{1/2d}\,e^{-3(s_{\rm in}-s)/2da_0}.
\end{equation}
Hence, initially, by which we mean at a radius $R_{\rm in}\gg1$ (e.g. $R_{\rm in}\sim (\ln S)^{1/d}$), $\Gamma_{\rm in}/H_{\rm in}\sim e^{3\phi_{\rm in}}s_{\rm in}^{1/2}$. So, in the large radius regime where the number density of long strings is negligible compared to the number density of small loops, $(S\gg) R^d\gtrsim\ln S$, we find that the change in radius during which $\Gamma/H\gg1$ will be
\begin{equation}\label{eq: DR/R int}
\frac{R-R_{\rm in}}{R_{\rm in}}\lesssim g_{\rm in}^6s_{\rm in}e^{\frac{3}{d}(s_{\rm in}/a_0)}-1.
\end{equation}
We then find that to have a positive increase in radius with the ensemble in thermal equilibrium, $\Gamma/H\gg1$, requires the initial string coupling to satisfy
\begin{equation}\label{eq: g_constr}
g_{\rm in}\gtrsim \frac{e^{s_{\rm in}/2da_0}}{s_{\rm in}^{1/6}}.
\end{equation}
Therefore, from (\ref{eq: g_constr}) we learn that maintaining thermal equilibrium seems to require a strongly coupled initial configuration.

Let us now turn to the issue of strong gravity and Jeans instabilities. Avoiding Jeans instabilities is probably not a fundamental requirement: It merely must be satisfied to ensure that a thermodynamic treatment is justified. Jeans instabilities occur when gravity becomes strongly coupled but we do not know the law of gravity at strong coupling (e.g. there may be some non-perturbative strong coupling effect which prevents the formation of Jeans instabilities). Having said that, if we do require that there be no Jeans instabilities then the dilaton is constrained \cite{{Atick:1988si}} by $e^{2\phi}\rho R^2\ll1$. We also require $R\gg1$ for the above formalism to apply and so these two conditions can be satisfied provided $e^{2\phi}\rho\ll1$. We can hence constrain the dilaton and the entropy density according to
\begin{equation}\label{eq: ephi constr}
e^{2\phi}\ll s^{-1}.
\end{equation}
We are to distrust the outcome of our theory if this condition is violated. To be specific, violation of (\ref{eq: ephi constr}) does not imply that thermodynamic equilibrium cannot be maintained. This can also be viewed as an initial condition constraint. As we have shown above both the dilaton and density decrease with increasing volume so their maximum values will in fact be their initial values. So, from (\ref{eq: intrate}) and (\ref{eq: ephi constr}) we find that
\begin{equation}\label{eq: intrate_jeans}
\frac{\Gamma}{H}\ll s_{\rm in}^{-3/2}\left(\frac{s_{\rm in}}{a_0}\right)^{1/2}\left(\frac{s}{s_{\rm in}}\right)^{1/2d}\,e^{-3(s_{\rm in}-s)/2da_0}.
\end{equation}
However, $s\leq s_{\rm in}$ {and so $\Gamma/H\ll s_{\rm in}^{-1}\ll1$}. Therefore, in the domain of validity of the above formalism the string ensemble cannot be in thermal equilibrium; we have however seen some hints that an initially strongly coupled dilaton may have an important role to play.

%%%%%%%%%%%%%%%%%%%%%%%%%%%%%%%%%%%%%%%%%%%%%%%%%%%%%%%%%%%%%%
\subsection{The Size Problem}
At present we know from measurements of the CMB power spectrum at large angular scales that if the Universe is toroidal, its radius, $\tilde{R}_0$, cannot be much smaller than the Hubble length today, $\tilde{H}_0^{-1}$, (see e.g. \cite{Levin:2001fg}): 
\ben\label{eq: obs}
\tilde{R}_0\tilde{H}_0=R_0H_0 \ge 1,
\een
where we have used the fact that the string frame ratio $RH$ is equal to the Einstein frame ratio $\tilde{R}\tilde{H}$ when the dilaton is constant. Denoting the radius at the end of the Hagedorn phase and the beginning of the radiation era by $R_r$ (and $\tilde{R}_r$ in the Einstein frame), and assuming adiabatic expansion, we find
\ben\label{eq: size}
\tilde{R}_r \gtrsim \left( \frac{\tilde{T}_0}{\tilde{T}_r}\right) \tilde{R}_0 \gtrsim 10^{-2}\;\mathrm{mm}
\een
where $T_r$ and $T_0$ are the temperatures at the start of the radiation era and today respectively, with $\tilde{T}_r \lesssim \mpl$. Hence the universe at the end of the Hagedorn era is very large in string units, justifying our concentrating on the large-radius thermodynamics.

The fact that the universe is large can be re-expressed as problem for the dilaton, as follows.
Working in the Einstein frame and taking the number of large dimensions $d=3$, we can estimate the ratio, $\tilde{R}_r\tilde{H}_r$, at time $\tilde{t}_r$, when the FRW radiation era (constant dilaton) begins, given that
\begin{equation*}
\tilde{H}^2(\tilde{t}) \simeq \tilde{H}^2_0 \left(\frac{\tilde{R}_0}{\tilde{R}_{\rm eq}}\right)^3 \left(\frac{\tilde{R}_{\rm eq}}{\tilde{R}(\tilde{t})}\right)^4,
\end{equation*}
where $\tilde{R}_{\rm eq}$ is the radius at matter-radiation equality. Note also that we are neglecting the dark energy era for our rough estimate. It then follows that
\begin{eqnarray*}
\tilde{R}_r\tilde{H}_r & \simeq & \tilde{R}_0\tilde{H}_0 \left(\frac{\tilde{R}_r}{\tilde{R}_0}\right) \left(\frac{\tilde{R}_0}{\tilde{R}_{\rm eq}}\right)^\frac32 \left(\frac{\tilde{R}_{\rm eq}}{\tilde{R}_r}\right)^2 \nonumber\\
&=& \tilde{R}_0\tilde{H}_0  \left(\frac{\tilde{R}_{\rm eq}}{\tilde{R}_0}\right)^\frac12\left(\frac{\tilde{R}_0}{\tilde{R}_r}\right).
\end{eqnarray*}
Given $\tilde{R}_{\rm eq}/\tilde{R}_0 \simeq 10^{-4}$, and that the entropy density $\tilde{s} \propto \tilde{R}^{-3}$ in the FRW cosmology, we have
\begin{equation*}
\tilde{R}_r\tilde{H}_r\ge 10^{-2} \left(\frac{s_r}{s_0}\right)^\frac13 \simeq 10^{-2} \left(\frac{\tilde{T}_r}{\tilde{T}_0}\right)
\end{equation*}
Therefore, at the end of the Hagedorn regime this ratio should satisfy
\begin{equation}\label{eq: size}
R_rH_r =\tilde{R}_r\tilde{H}_r\ge 10^{29} \left(\frac{\tilde{T}_r}{\mpl}\right),
\end{equation}
assuming $\tilde{T}_r \sim \tilde{T}_{\rm Hag} \simeq \mpl$. Hence, at the beginning of the radiation era, independently of whether we are working in the string or Einstein frame, the Hubble length must occupy a tiny fraction of the size of the Universe for a toroidal background to be phenomenologically acceptable.  

We can estimate the ratio $R_rH_r$ on account of the string frame expression (\ref{eq: Hsolapx}) with $w=a_0/s$, $S_H=sH^{-d}$ and $S=sR^d$, given that at the onset of radiation $s\sim1$,
\begin{equation}\label{eq: RrHr}
R_rH_r\simeq s_{\rm in}^{3/(2d)}\exp\left(-\frac{s_{\rm in}}{2da_0}\right)R_{\rm in}H_{\rm in}.
\end{equation}
Therefore, one way of satisfying the constraint (\ref{eq: size}) is for the initial (string frame) Hubble parameter,
\begin{equation*}
H_{\rm in} \ge 10^{29} \left(\frac{\tilde{T}_r}{\mpl}\right)R_{\rm in}^{-1}s_{\rm in}^{-1/2}e^{s_{\rm in}/(6a_0)},
\end{equation*}
where we have taken $d=3$. For temperatures $T_r\sim \mpl$ and given that $s_{\rm in}\gg1$, this is much larger than that allowed from the assumption of adiabaticity according to which $|H|<1$. On account of the initial conditions constraint (\ref{eq: H0}) one sees that this inequality can be rewritten in terms of the initial string coupling, $g_{\rm in}$,
\begin{equation*}
g_{\rm in}\gtrsim 10^{29}\left(\frac{\tilde{T}_r}{m_{\rm Pl}}\right)R_{\rm in}^{-1}e^{s_{\rm in}/(6a_0)},
\end{equation*}
which is inconsistent with requirement of weak coupling. We conclude that the 3-torus (on which our universe is embedded in this scenario) cannot become large enough to agree with observations, namely equation (\ref{eq: obs}), if string is weakly coupled close to the string scale, $g_{\rm in}\ll1$.

%%%%%%%%%%%%%%%%%%%%%%%%%%%%%%%%%%%%%%%%%%%%%%%%%%%%%%%%%%%%%
\subsection{The Dilaton Problem}
A dilaton problem arises in the following way. We have seen that the onset of radiation domination is characterised by a string frame entropy density of order one, $s_r=S/R_r^3\sim1$. We can derive the entropy density in the Einstein frame, $\tilde{s}=S/\tilde{R}^3$, on account of $\tilde{R}=g^{-1}R$ which relates Einstein and string frame radii (\ref{eq: G_E vs G}). We then find that the Einstein frame entropy density at the onset of radiation is
\begin{equation}\label{eq: sE_gr}
\tilde{s}_r\sim g_r^3. 
\end{equation}
The entropy density is approximately $\tilde{s} \sim \tilde{T}^3$ and hence
\begin{equation*}
g_r\sim \frac{T_r}{m_{\rm Pl}},
\end{equation*}
where we have explicitly restored the units. 
To a first approximation one expects the temperature at the end of the Hagedorn region to be $T_r\sim\mathcal{O}(m_{\rm Pl})$ and hence we have shown that the string coupling should be of order one,
\begin{equation}\label{eq: g_r p}
g_r\sim1.
\end{equation}
This we expect independently from the relation between gauge and string couplings in a universe where the compact dimensions remain at the string scale. However, as we have seen above, during the Hagedorn era the dilaton is expected to run to negligible coupling and at the onset of radiation (\ref{eq: gr}),
\begin{equation}\label{eq: gs_r}
g_r\sim g_{\rm in}e^{-s_{\rm in}}\ll1,
\end{equation}
where $g_{\rm in}\lesssim 1$. It is hence clear that there is large disagreement between the string coupling at the onset of radiation domination as required from the post-Hagedorn considerations (\ref{eq: g_r p}) and that predicted from the equations of motion during the large radius Hagedorn regime (\ref{eq: gs_r}).

%%%%%%%%%%%%%%%%%%%%%%%%%%%%%%%%%%%%%%%%%%%%%%%%%%%%%%%%%%%%%%%%%
\section{Discussion}
We have considered an ensemble of high energy strings in dilaton gravity background. The picture we have is the following: the universe is initially at string scale, all dimensions are of equal size and all degrees of freedom are present. Quantum fluctuations create a hierarchy of scales and based on the original formulation \cite{Brandenberger:1988aj} 3 dimensions will become large. This is our starting point for the calculations presented in this paper, 3 dimensions are large and 6 remain at string scale. 
%The stabilisation mechanism of the radions was originally based on the fact that winding modes are confining (see \cite{Brandenberger:1988aj} and also \cite{Tseytlin:1991xk} for further details) but in this paper we have assumed that they are stabilised. 

We have calculated the equation of state for a (heterotic) string gas and have used it to determine the resulting dynamics of the metric and dilaton fields. We have concentrated on the large radius Hagedorn regime with totally compact topology and have allowed $d$ of the $9-d$ spatial dimensions to freely evolve. We have found analytic solutions for both the Hubble parameter and dilaton as functions of size for the entire large radius Hagedorn evolution and have compared our results with numerical solutions. %In deriving the equation of state we have also corrected a small error in the standard calculation of the microcanonical density of states (\ref{eq: Omega}) calculated in \cite{Deo:1991mp}. 
We have found that the equation of state has the form 
$$
w(x)\simeq \frac{x}{1-x},
$$ 
with $x/a_0$ the inverse entropy density, $s$, and $a_0$ a constant arising due to the presence of momentum modes. Higher order terms can be found in (\ref{eq: alpha(lambda)}). For large entropy densities, $s$, we have found that $\phi(s)\simeq \phi_{\rm in}-(s_{\rm in}-s)/2da_0,$ and $H(s)\simeq H_{\rm in}(s/s_{\rm in})^{1/2d}\exp\{-(s_{\rm in}-s)/2da_0\}$, the full expressions are given in (\ref{eq: dil_nonad}) and (\ref{eq: Hubble_nonad}) respectively. In particular, the dilaton is shown to decrease inversely proportional to the volume of the large compact dimensions. The Hubble parameter decreases with increasing volume, reaches a minimum when the equation of state $w\sim 1/d$ and then may increase slightly, this is sensitive to the dependence of the total density of states, $\Omega$, on $E$ and $R$ when the energy density $\rho$ is close to the Hagedorn energy density $\rho_1$. Integrating the leading order form for the Hubble parameter as a function of size we have derived the leading dependence of size with time (\ref{eq: xt}) and have obtained an estimate for the duration of the Hagedorn era in the large radius Hagedorn regime (\ref{eq: t_H}), $$\Delta t_H\sim \frac{1}{g_{\rm in}}\frac{e^{s_{\rm in}}}{\sqrt{s_{\rm in}}},$$
which is very large if the dilaton is weakly coupled. 

It is also useful to compare our results for radius and dilaton as functions of time for large radius evolution with the corresponding results for vanishing pressure \cite{Tseytlin:1991xk} at small radius. A main difference between small and large radius evolution is that at small radii the pressure is very small and slowly evolving while at large radii the pressure is approximately constant and of $\mathcal{O}(1)$. It suffices to say that at large times TV \cite{Tseytlin:1991xk} find that $\phi(t)\sim-\ln t+{\rm const}$. Our solution (\ref{eq: phit}) agrees in that we have also found a logarithmic dependence of dilaton on time, $\phi(t)\sim -\ln(1+g_{\rm in}\sqrt{s_{\rm in}}\Delta t/2)+{\rm const}$. As far as the radius is concerned we have found that there is no maximum radius due to the fact that the equations of state increases with increasing size. In particular, TV \cite{Tseytlin:1991xk} find that $\lambda=\lambda_{\rm in}+\ln(\frac{t-c}{t+c})^{1/\sqrt{d}}$ for some integration constant $c$ whereas we find that in the large radius region $\lambda\simeq \lambda_{\rm in}+\ln[1-\gamma\ln(1+H_{\rm in}\Delta t/\gamma)]^{-1/d}$ with $\gamma\equiv 2da_0/s_{\rm in}\ll1$ valid for $\Delta t\lesssim e^{s_{\rm in}}/(g_{\rm in}\sqrt{s_{\rm in}})$ (use has also been made of (\ref{eq: H0})).

Therefore, the results we have obtained differ from the small radius evolution. The source of these differences boil down to the string spectrum. In particular, even though the oscillator modes account for most of the energy of the ensemble, the number of long strings at large radii, $S^{1/d}\gg R\gg(\ln S)^{1/d}$, is negligible compared to the number of small loops present. The two types of string contribute equally to the number density when the radius $R\sim(\ln S)^{1/d}$. Furthermore, the massless modes are expected to dominate the ensemble below the Hagedorn energy density \cite{Brandenberger:1988aj} where $R\sim S^{1/d}$. If above the radius $R\sim(\ln S)^{1/d}$ winding strings are few in number then one would naively expect that the splitting from 9 to 6+3 dimensions (in the original manifestation of the Brandenberger-Vafa mechanism of spacetime dimensionality) should occur before this radius is reached. However, as we have stressed above evolution in the small radius regime is very speculative to date. The main reason being that one expects there to be non-perturbative effects that come into play and these may in turn give rise to a vastly different picture from the one obtained by perturbative calculations alone. 

In the last section we have seen that there are a number of problems associated with a large radius high-entropy string gas universe, associated with the lack of thermodynamic equiilibrium and the requirement to have a large but not too weakly-coupled universe today. One cannot rely on early strongly-coupled dynamics to retrieve the scenario, as 
a small radius universe must be followed by a large radius one, with large entropy if we are to describe our universe, and our analysis applies to this phase.  This version of string gas cosmology therefore still has many difficulties to overcome.

%%%%%%%%%%%%%%%%%%%%%%%%%%%%%%%%%%%%%%%%%%%%%%%%%%%%%%%%%%%%%%%%%%%%%%

\begin{acknowledgements}
DPS would like to thank R. Brandenberger, A. Linde, Chung-I Tan, A. Mazumdar, E. Alvarez, D. Bailin, S. Huber, and S. Giddings for various help during the completion of this paper. DPS is supported by STFC.
\end{acknowledgements}

%%%%%%%%%%%%%%%%%%%%%%%%%%%%%%%%%%%%%%%%%%%%%%%%%%%%%%%%%%%%%
\nocite{*}
\bibliographystyle{apsrev}
\bibliography{sg}

\end{document}